\documentclass[prd,reprint,onecolumn,preprintnumbers,unsortedaddress,superscriptaddress,notitlepage,showpacs]{revtex4}
\usepackage{amsfonts}
\usepackage{amsmath}
\usepackage{hyperref}
\usepackage{amssymb}
\usepackage[english]{babel}
\usepackage{graphicx}
\usepackage{epsfig}
\usepackage{bm}
\usepackage{longtable}
\usepackage{verbatim}
\usepackage{longtable}
\usepackage{amsfonts}
\usepackage{amsthm}
\usepackage{graphicx}
\usepackage{axodraw4j}

\newcommand{\n}{\~n}

\newcommand{\beq}{\begin{equation}}
\newcommand{\eeq}[1]{\label{#1}\end{equation}}
\newcommand{\bea}{\begin{eqnarray}}
\newcommand{\eea}[1]{\label{#1}\end{eqnarray}}

\begin{document}

\title{A SUSY $SU(6)$ GUT model with Pseudo-Goldstone Higgs Doublets\\
Un Modelo Supersim\'etrico $SU(6)$ de Gran Unificaci\'on con Doubletes de Higgs Pseudo-Goldstone}

\author{A.E. C\'arcamo Hern\'andez}
\email{antonio.carcamo@usm.cl}
\affiliation{Universidad T\'ecnica Federico Santa Mar\'{\i}a and Centro Cient\'{\i}fico-Tecnol\'{o}gico de Valpara\'{\i}so. Avda. Espa\n a 1680 Casilla 110-V,\\ Valpara\'{\i}so, Chile}
\author{Rakibur Rahman}
\email{rakibur.rahman@aei.mpg.de}
\affiliation{Max-Planck-Institut fur Gravitationsphysik (Albert-Einstein-Institut), \\
Am Muhlenberg 1, D-14476 Potsdam-Golm, Germany}

\begin{abstract}
We present a novel way of realizing the pseudo-Nambu-Goldstone boson
mechanism at all orders in perturbation theory, for the doublet-triplet splitting in supersymmetric grand unified
theories. The global symmetries of the Higgs sector are attributed to a
non-vectorlike Higgs content, which is consistent with unbroken supersymmetry
in a scenario with flat extra dimensions and branes. We also show how in such
a model one can naturally obtain a realistic pattern for the Standard Model
fermion masses and mixings.\\
Presentamos un modelo que genera el mecanismo de pseudo-Nambu-Goldstone a todo orden en teor\'{i}a de perturbaciones para el problema de corrimiento de doblete-triplete en teor\'{\i}as supersim\'etricas de gran unificaci\'on. Las simetr\'{\i}as globales del sector de Higgs son atribu\'{\i}das al contenido no vectorial de Higgs, el cual es consistente con supersimetr\'{\i}a no rota en un escenario de dimensiones extra planas y branas. Adem\'as, mostramos como en este modelo uno puede obtener un patr\'on real\'{\i}stico de las masas y mezclas de fermiones del Modelo Est\'andar.
\end{abstract}


\maketitle

\section{Introduction}
The discovery of the Higgs Boson at the LHC \cite{atlashiggs,cmshiggs} by ATLAS and CMS collaborations, has concreted the great success of the Standard Model (SM) in
describing electroweak phenomena. Nonetheless, the SM does not explain neither
the fermion mass and mixing pattern nor the existence of three fermion families. Furthermore, the SM has the hierarchy problem caused by the quadratic divergence of the Higgs mass, which gives an indication of an unknown underlying physics in the gauge symmetry breaking mechanism. Consequently, a more fundamental theory is needed to address these issues. The existing fermion mass pattern spreads over a range of five orders of magnitude in the quark
sector and a dramatically broader range in the neutrino sector. The experimental well established fact that in the quark sector the mixing angles are small, wheareas two of the leptonic mixing angles are large, and one is small; suggests that the corresponding mechanisms for masses and mixings should be different.

Models with an extended gauge symmetry are frequently used to address the problems of the SM. In particular, grand unified theories (GUTs) are a major attempt beyond the standard model (SM) to provide a unified gauge theoretic description. The scale at which
GUTs must replace the SM, however, should be very high, $M_\text{G}> 10^{15-16}$
GeV, in order to suppress higher dimensional operators that would otherwise
lead to a large proton decay rate. Again, in the minimal supersymmetric
standard model (MSSM) the different gauge couplings do unify at $M_\text{G}\sim
10^{16}$ GeV. Because supersymmetry (SUSY) also helps with the stability of the
weak scale against the large GUT scale, supersymmetric GUTs provide a very
appealing framework for physics beyond the SM.

The most problematic aspect of supersymmetric GUTs is the doublet-triplet
splitting (DTS) problem. The minimal $SU(5)$ supersymmetric GUT, for example,
knows only about very large scales: $M_\text{G} \sim 10^{16}$ GeV and $M_\text P
\sim 10^{18}$ GeV. Yet it should somehow give rise to a pair of essentially
massless electroweak Higgs doublets $h, \bar h$ that survive down to the low
energies, and are not accompanied by the color triplets. While the Higgs doublet
masses should be around 100 GeV, the triplet masses are around $10^{16}$
GeV, as seen from higgsino-mediated proton decay calculation in the simplest
models. It is difficult to understand how fields from a single GUT representation
can have such different masses. In other words, supersymmetric GUTs involve a
parameter with an accuracy of $\mathcal{O}(10^{-13})$! One may wonder how
acceptable a model is with a parameter as small as $10^{-13}$.

To avoid this fine tuning and naturally distinguish the doublet and triplet
Higgs masses, a number of solutions have been proposed~\cite{missingpartner,
missingvev,g1,g2,u5,bd,bdm,bcr}. Particularly appealing among them are the models
with Higgses as pseudo-Nambu-Goldstone bosons (PNGBs)~\cite{u5,bd,bdm,bcr,GBMandLH}. For some works having the implementation of the Goldstone boson mechanism to solve the little Hierarchy problem, see Refs. \cite{GBMandLH}. The idea is that one can identify the Higgs doublets as the zero modes of a
compact vacuum degeneracy, rendered massless by supersymmetry to all orders in
perturbation theory. Once (soft) SUSY breaking terms are included, the flat
directions are lifted and the doublets acquire masses of just the right order
of magnitude. This distinguishes the doublets
from the triplets in a nontrivial way, so that it is natural to obtain light
doublets while the remaining fields are heavy. The model of~\cite{bd,bdm},
which we will briefly discuss below, is a nice realization of this idea.

The key observation in \cite{bd,bdm} is that a compact degeneracy,
giving automatically heavy color-triplet partners that decouple along the
flat directions, is possible if the different Higgs fields that break the
grand unified gauge group have no cross-couplings in the superpotential.
Thanks to the independent global rotations of the uncorrelated vacuum
expectation values (VEVs) that give an accidental flat direction to the
vacuum. However, this rotation $-$ broken by the gauge and Yukawa couplings
$-$ is not an exact symmetry of the theory, so that the corresponding zero
modes are physical, for they are not eaten up by any gauge field. The model
is based on $SU(6)$ gauge group with minimal supersymmetry. Breaking the
symmetry down to the standard model group, $G_\text{SM} = SU(3)_c\otimes
SU(2)_L\otimes U(1)_Y$, requires at least two Higgs representations: an
adjoint $\Sigma_\alpha^\beta$ and a fundamental--antifundamental pair $%
\Phi_\alpha, \bar{\Phi}^\alpha$ ($\alpha,\beta = 1,2,...,6$). The Higgses
develop the following VEVs: \beq \langle\Sigma_\alpha^\beta\rangle~=~{\text{diag}}
(1,1,1,1,-2,-2)\,\sigma~,\qquad \langle\Phi_\alpha\rangle~=~\langle\bar{\Phi}^\alpha
\rangle~=~(\phi,0,0,0,0,0)~.\eeq{r1} While the first VEV leaves unbroken
$G_{\Sigma} = SU(4)_c\otimes SU(2)_L\otimes U(1)$, the second one preserves
$G_{\Phi} = SU(5)$, so as to give unbroken $G_\text{SM}$ as the intersection.
Now if these two sectors have no cross-couplings in the superpotential,
\beq W~=~W(\Sigma) + W(\Phi,\bar{\Phi})~, \eeq{r2}
there is an effective global symmetry $G_\text{gl} = SU(6)_{\Sigma}\otimes
SU(6)_{\Phi}$. For the VEVs given in~(\ref{r1}) this global symmetry is broken
to $G_{\Sigma}\otimes G_{\Phi}$, and the vacuum will have compact flat
directions, which do not pertain to any broken generator of the gauge group.
It is easy to count
the number of the Goldstone modes and of the broken gauge generators; the
former exceeds the latter by two. The following linear combinations of the
electroweak doublets (coming from the $\Sigma$ and $\Phi, \bar{\Phi}$ fields)
are the two left-over zero modes: \beq h~=~{\frac{\phi h_{\Sigma} - 3\sigma
h_{\Phi}}{\sqrt{\phi^2 + 9\sigma^2}}}~,\qquad \bar h~=~{\frac{\phi\bar
h_{\Sigma} - 3\sigma\bar h_{\bar{\Phi}}}{\sqrt{\phi^2 + 9\sigma^2}}}~.\eeq{r3}
The DTS problem is solved, for all other states are heavy. We note that in order
to get correct order of symmetry breaking and successful prediction of
$\sin^2\theta_\text W$, one needs to have $\phi>\sigma\sim M_\text{G}$~\cite{bdm, bcr, bdsbh}.
The light Higgses are therefore predominantly contained in $\Sigma$.

The main problem of this model is to justify the absence of the cross-coupling
$\Phi\Sigma \bar{\Phi}$, which is not forbidden by the gauge symmetry.
Existence of such a coupling would break the $SU(6)_{\Sigma}\otimes
SU(6)_{\Phi}$ global symmetry of the Higgs sector, thereby destroying the
PNGB mechanism for the light Higgs doublets. One may invoke some extra
discrete symmetries or a larger gauge symmetry for this coupling to be
absent~\cite{bdm,bcr,Be}. Moreover, because of quantum gravity effects,
Planck-suppressed higher dimensional operators compatible with the symmetries
may show up with $\mathcal{O}(1)$ coefficients. As $M_\text{G}$ is not very small
compared to $M_\text P$, one must forbid the cross-couplings to very high orders,
which again may require some unappealing symmetries or charge assignments.

Obtaining a realistic pattern for the SM fermion masses and mixings is
another problem. The smallest anomaly-free set of chiral representations
of $SU(6)$ is $\mathbf{15+\bar{6}+\bar{6}}$, which can contain a family of
light matter fields. If the top quark is contained in a $\mathbf{20}$
(pseudo-real representation) of $SU(6)$, then the interaction $\mathbf{20}\,
\Sigma \,\mathbf{20}$ exclusively gives the top quark an $\mathcal{O}(1)$
Yukawa coupling. Masses of other fermions arise from Planck-suppressed
non-renormalizable operators. However, it is hard to obtain a realistic fermion
mass pattern if one incorporates all the possible non-renormalizable operators.
One needs to appeal to extra discrete symmetries, and assume that the higher
dimensional operators come from integrating out some heavy vector-like
fields~\cite{bdsbh,Be}.

The purpose of this paper is to present a supersymmetric GUT model which
naturally circumvents most of the aforementioned problems. In the next section
we consider anomaly-free combinations of \emph{chiral} Higgses, which help
with the global $SU(6)\otimes SU(6)$ symmetry of the Higgs sector, as was first
noted in~\cite{u1A}. We choose a Higgs content that allows vacua with unbroken
SUSY under certain conditions. In Section 3 we propose a model with a flat
extra dimension and branes, where the chiral Higgs multiplets are localized on
two separate branes, which not only makes the global symmetry automatic, but also
ensures the existence of a SUSY-preserving vacuum. We further extend the model,
and specify the Higgs VEVs to be able to produce naturally, without appealing to
flavor symmetries, a realistic pattern for fermion masses and mixings, which we
work out in Section 4. Finally in Section 5 we make some concluding remarks.

\section*{2. Non-Vectorlike Higgses}

Our starting point is the fact, exploited in~\cite{u1A}, that certain cross-couplings
in the superpotential are automatically absent if the Higgs content has non-vectorlike
representations of the gauge group. We would like to consider a chiral anomaly-free
combination of Higgs multiplets. An adjoint of $SU(6)$, $\Sigma$, does not contribute
to anomaly. One can replace the Higgses $\mathbf 6, \mathbf{\bar 6}$, considered in~\cite{bd,bdm},
by an anomaly-free non-vectorlike set of Higgses.

It is easy to see that $\mathbf{15+\bar{6}+\bar{6}}$ does \emph{not} work. We
want to give VEVs to this set in such a way that the $SU(6)$ gauge group breaks
down to $SU(5)$. Then the $\mathbf{35}$ will break it down to $G_\text{SM}$ in
the usual way. Thus the $\mathbf{\bar{6}}$'s, which we will denote as $\bar\Phi_i^\alpha$,
can only acquire VEVs like $(\phi, 0, 0, 0, 0, 0)$. On the other hand, $\mathbf{15}(\Theta)$
being an antisymmetric representation, can only have off-diagonal nonzero elements.
Since we want to have unbroken SUSY, the VEVs should give zero scalar
potential: both $D$-terms and $F$-terms must vanish separately. Now the $D$%
-term contribution from $\Sigma$ is automatically zero. Considering the VEVs
$\langle\bar{\Phi}_i^\alpha\rangle = \phi_i\delta_1^\alpha$, and say
$\langle\Theta_{\alpha\beta}\rangle=\theta(\delta_\alpha^1\delta_\beta^2-
\delta_\alpha^2\delta_\beta^1)$, we have
\bea
\langle D^a_{\bar{\Phi}}\rangle&=&\langle\bar{\Phi}^ *_{i\alpha}\{-\bar{\Phi}%
^\beta_i(T^a)_\beta^\alpha\}\rangle~=~-\sum_i\phi_i^2(T^a)_1^1~,\label{r06}\\
\langle D^a_{\Theta}\rangle &=&
\langle\Theta^{*\alpha\beta}\{\delta_\alpha^{\alpha^\prime}(T^a)_\beta^
{\beta^\prime}+(T^a)_\alpha^{\alpha^\prime}\delta_\beta^{\beta^\prime}\}
\Theta_{\alpha^\prime\beta^\prime}\rangle~=~2\theta^2\{(T^a)_1^1 +(T^a)_2^2\}~.
\eea{r061}
These cannot cancel in general for nonzero VEVs. Therefore the $D$-terms do
not vanish.

On the other hand, we can consider the set $\mathbf{21+\bar{6}+...+\bar{6}}$,
with 10 $\mathbf{\bar{6}}$'s. Let us denote the $\mathbf{21}$ as $\Psi_{\alpha\beta}$.
With the VEV: $\langle\Psi_{\alpha\beta}\rangle=\psi\delta_\alpha^1\delta_\beta^1$,
we have \beq
\langle D^a_{\Psi}\rangle~=~\langle\Psi^{*\alpha\beta}
\{\delta_\alpha^{\alpha^\prime}(T^a)_\beta^{\beta^\prime}
+(T^a)_\alpha^{\alpha^\prime}\delta_\beta^{\beta^\prime}\}
\Psi_{\alpha^\prime\beta^\prime}\rangle~=~2\psi^2(T^a)_1^1~.
\eeq{r6}
In view of~(\ref{r06}) it is thus always possible to have vanishing contribution
coming from the $D$-terms by suitably choosing $\psi$:
\beq \psi~=~\frac{1}{\sqrt{2}}\,\left(\sum_{i=1}^{10}\phi_i^2\right)^{1/2}.\eeq{r7}
Next, in order to consider the $F$-terms, we write the superpotential
\bea W&=&W(\Sigma)+W(\Psi,\bar{\Phi})  \notag \\
\notag \\
&=&\left\{\frac{M}{2}\text{Tr}\Sigma^2+ \frac{\lambda}{3}\text{Tr}%
\Sigma^3\right\}+\sum_{i,j}g_{ij}\Psi_{\alpha \beta}\bar{\Phi}_i^\alpha\bar{%
\Phi}_j^\beta~.\eea{r8} Note first that we could have included in $W(\Sigma)$
higher dimensional operators. These could only modify slightly the magnitude
and orientation of a SUSY-preserving VEV $\langle\Sigma_\alpha^\beta\rangle$,
and are not important for us. Second, cross-couplings of the form $W(\Sigma,\Psi)$
and $W(\Sigma,\bar{\Phi})$, which could otherwise destroy the global symmetry
$SU(6)\otimes SU(6)$ of the superpotential, are automatically forbidden by the
gauge symmetry. Of course, one could still have bad cross-couplings of the kind
$W(\Sigma,\Psi,\bar{\Phi})$ at the non-renormalizable level. These however will
also be absent, thanks to the setup that we will consider in the next section.

It is well-known that the VEV $\langle\Sigma_\alpha^\beta\rangle = {\text{%
diag}}(1,1,1,1,-2,-2)\,\sigma$, where $\sigma=M/\lambda$, gives vanishing $F$%
-term for $\Sigma$. We require that the other $F$-terms also vanish:
\bea
\langle F_{\bar{\Phi}}\rangle&\equiv&\left\langle \frac{\partial W}{\partial
\bar{\Phi}_i^\alpha}\right\rangle~=~2\sum_jg_{ij}\langle\Psi_{\alpha\beta}%
\bar{\Phi}_j^\beta\rangle~=~0,\label{r9} \\
\langle F_{\Psi}\rangle &\equiv& \left\langle\frac{\partial W} {%
\partial\Psi_{\alpha\beta}}\right\rangle~=~\sum_{i,j}g_{ij} \langle\bar{\Phi}%
_i^\alpha\bar{\Phi}_j^\beta\rangle~=~0. \eea{r10}
With the given VEVs, the above requirements are nontrivial only when $%
\alpha=\beta=1$. In the latter case, it is necessary and sufficient to
require \beq \sum_jg_{ij}\phi_j~=~0. \eeq{r11}
That is, the $10\times10$ coupling matrix $g_{ij}$ has (at least) one zero
eigenvalue, with $\phi_i$ being the corresponding eigenvector. Therefore if
$\det g_{ij}=0$, SUSY-preserving vacua exist.

One can work out the light Higgs doublets; they are linear combinations of the
electroweak doublets coming from the $\Sigma$ and $\Psi, \bar{\Phi}_i$ fields:
\beq h~=~\frac{h_{\Sigma}}{\sqrt{1+\kappa^2}}- \frac{\kappa h_{\Psi}}{\sqrt{1+\kappa^2}}~,
\qquad\bar{h}~=~\frac{\bar{h}_{\Sigma}}{\sqrt{1+\kappa^2}}-\frac{\kappa}{\sqrt{1+\kappa^2}}
\sum_{i=1}^{10}\tfrac{1}{\sqrt{2}\psi}\,\phi_i\bar{h}_{\bar{\Phi}_i},\eeq{r12} where $\kappa$
is given by \beq \kappa~\equiv~\frac{3\sigma}{2\psi}~.\eeq{r13} As mentioned in the
introduction, the $SU(6)$ breaking scale ($\sim\psi$) should be larger than the breaking
scale of $SU(5)$, which is $\sigma\sim M_\text{G}\sim 10^{16}$ GeV, i.e. $\psi\gtrsim\sigma$.
Then the ratio $\kappa$ is less than unity, so that the light Higgses are predominantly
contained in $\Sigma$.

\section*{3. The Model}

Our model consists of two parallel 3-branes separated by a distance $d$ in a flat
$(4+1)$D bulk space-time. The single extra dimension\footnote{Just as in the model
of~\cite{Cheng}, one extra dimension turns out to be favorable in order to obtain
successful fermion masses. Our model is different, though, from that
of~\cite{Cheng}, most importantly, in the Higgs content. Namely, the Higgs content
of the latter is $\mathbf{35}+\mathbf{6}+\mathbf{\bar{6}}$, which is vector-like.}
is compactified with a radius $r$, larger than the 4D Planck length $M_\text P^{-1}$,
so that the fundamental scale of quantum gravity, $M_*$, is lower than $M_\text P$~\cite{add}.
However, we still take $r$ to be so small as to have $M_*>M_\text{G}\sim 10^{16}$ GeV.
This ensures that the gauge coupling unification works successfully as usual
(discussions of the gauge coupling unification issue can be found in~\cite{ddgross}).
While $d<r$, the 5D Planck length $M_*^{-1}$ is still assumed to be much smaller
than $d$; this enables us to describe physics by the usual field theory language,
without caring about quantum gravity effects.

In this setup we have the $SU(6)$ gauge field, the Higgs fields, various matter
fields, and some vector-like heavy fields $-$ some living in the bulk, some
confined in one of the branes. The extra dimension is assumed to be compactified
on an orbifold, so that we can get chiral multiplets in four dimensions, with
unwanted zero modes projected out. By integrating out the extra dimension, one
obtains an effective 4D Lagrangian, which makes sense at low energies. Among
others, this Lagrangian contains light fields, that may come from either of the
branes and the bulk. In the 5D setup the various fields are localized in the
following way:
     \begin{itemize}
       \item The $SU(6)$ gauge field propagates in all bulk.
       \item The Higgses are localized on the branes: the $\mathbf{35}(\Sigma)$
and \emph{some}, but not all, of the 10 $\mathbf{\bar{6}}$'s on brane-1, and the
$\mathbf{21}(\Psi)$ and the \emph{other} $\mathbf{\bar{6}}$'s on brane-2, say.
We can always define the ones on brane-2 as $\{\bar{\Phi}_{i}^\alpha:~i=1,2,...,n<10\}$
by the $SU(10)$ global rotation of the $\mathbf{\bar{6}}$'s .
       \item There is a $\mathbf{20}$, living on the same brane as $\Sigma$,
that contains the top quark (and hence the top quark has $\mathcal{O}(1)$
Yukawa coupling). Let us call it $\xi$.
       \item All other matter fields and some additional heavy vector-like fields
live in the bulk.
     \end{itemize}

All Higgs fields $\Sigma$, $\Psi$, and $\bar{\Phi}_i$ are assumed to have a matter
parity $+1$, while $\xi$ and the other the matter fields and vector-like fields living
in the bulk have matter parity $-1$. This would help with the existence of vacua with
unbroken SUSY, and also forbid unwarranted non-renormalizable cross-couplings in the
superpotential.

To see how we have a SUSY-preserving vacuum, note that the 10 $\mathbf{\bar{6}}$'s are
split into two sets living on two spatially separated branes. Now let us consider a diagram
that can potentially generate an element $g_{ij}$ of the coupling matrix, with
$\mathbf{\bar{6}}_i$ and $\mathbf{\bar{6}}_j$ living on \emph{different} branes. Because of
the matter parity assignment, such a diagram cannot appear at tree level; it must contain at
least one loop with a bulk field running in it. Therefore, after integrating out the extra
dimension, we will have at least one zero eigenvalue for the coupling matrix, thanks to the
non-renormalization theorem.

The above reasoning also clarifies why potentially bad cross-couplings of the kind
$W(\Sigma,\Psi,\bar{\Phi})$, allowed by the gauge symmetry, will be absent; such couplings necessarily
involve Higgs fields from \emph{different} branes. Given this one justifies the form of the Higgs
superpotential~(\ref{r8}); the Higgs sector indeed has the global symmetry $SU(6)\otimes SU(6)$.

It is worth mentioning that in our model we are localizing on each brane a set of
fields that necessarily give rise to chiral anomaly on individual branes. However,
this per se is not an inconsistency; there will be a right amount of anomaly inflow
into each brane, since the full 5D theory is anomaly-free in the first place.

\subsubsection*{Further specification and extension for realistic model-building}

Without additional ingredients the above model, however, may not be able to produce
a realistic pattern for the fermion masses. One finds that by the
following specification and extension of the model, one can naturally obtain the SM
fermion masses and mixings. First, let us put on brane-2, along with $\Psi$, only 6
of the $\mathbf{\bar{6}}$-Higgses, say $(\bar{\Phi}_1,\bar{\Phi}_2,...,\bar{\Phi}_6)$.
The rest, namely $(\bar{\Phi}_7,\bar{\Phi}_8,\bar{\Phi}_9,\bar{\Phi}_{10})$, live on
brane-1, as does $\Sigma$.

We also extend our gauge group from $SU(6)$ to $SU(6)\otimes U(1)_\text A$, where the gauge
symmetry $U(1)_\text A$ is anomalous\footnote{Such an extension of $SU(6)$ supersymmetric GUTs
has been considered by the authors in~\cite{u1A}.}. Note that gauge anomalies are usually
present in string theory~\cite{anom}, and cancelation thereof takes place by the Green-Schwarz
mechanism~\cite{{GSM}}. It requires non-zero mixed anomalies, so that some of our fields
must be charged under $U(1)_\text A$. Such an anomalous gauge symmetry will always generate
a Fayet-Iliopoulos term~\cite{fet}, which is proportional to the sum of the charges.
Below we further specify the various fields, and the $U(1)_\text A$-charges thereof.

\begin{itemize}
  \item Let us have in the bulk three sets of matter multiplets: $(\zeta_i,\bar{\chi}_i,\bar{\chi}'_i)$,
with $i=1,2,3$, and the fields transforming respectively as $\mathbf{15}$, $\mathbf{\bar{6}}$,
and $\mathbf{\bar{6}}$ with respect to $SU(6)$. We assume that \emph{all} the matter fields
are neutral under the $U(1)_\text A$.
  \item Let there be several pairs of heavy vector-like fields of the $SU(6)$ representation:
$(\mathbf{6}_0,\mathbf{\bar{6}}_0)$, $(\mathbf{6}_{\pm1},\mathbf{\bar{6}}_{\pm1})$,
$(\mathbf{70}_0,\mathbf{\overline{70}}_0)$, $(\mathbf{70}_{\pm1},\mathbf{\overline{70}}_{\pm1})$,
$(\mathbf{20}_{\pm1},\mathbf{20}'_{\pm1})$, and $(\mathbf{20}_{\pm2},\mathbf{20}'_{\pm2})$,
with the subscripts denoting their respective charges under $U(1)_\text A$. For each pair, we assume
that the zero-mode masses do not depend on the $U(1)_\text A$-charge.
  \item For the Higgs sector, we assume the following charge assignment under $U(1)_\text A$:
\beq q_{\Sigma}~=~0,\hspace{0.25cm}q_{\Psi}~=~+1,\hspace{0.25cm}q_{\bar{\Phi}_i}~=~\left(-\tfrac{1}{2},-\tfrac{1}{2},
-\tfrac{1}{2},-\tfrac{1}{2},-\tfrac{1}{2},0,+1,+1,+1,-1\right).\eeq{f1}
\end{itemize}

It is noteworthy that the following nonrenormalizable term that could ruin the Goldstone boson mechanism, is invariant under the symmetries of the model:
\begin{equation}
\frac{1}{M^2_P}\text{Tr}\Sigma^2\sum_{i,j}g_{ij}\Psi_{\alpha \beta}\bar{\Phi}_i^\alpha\bar{%
\Phi}_j^\beta~.
\end{equation}
However that unwanted term does not appear in our model since the scalar fields $\Psi$ and $\Sigma$ are located at different branes. Consequently, the Goldstone boson mechanism holds at all orders in perturbation theory.

\subsubsection*{Choice of the Higgs VEVs}

Having assigned the charges as above, we see that the only non-zero elements our $10\times10$
coupling matrix $g_{ij}$ will have are in the upper left $5\times5$ block. This enables us to
have a SUSY-preserving VEV of the form \beq \phi_i~=~\phi\,(\,0,\,0,\,0,\,0,\,0,\,\epsilon,\,
\gamma,\,\gamma,\,\gamma,\,\epsilon'\,)~,\eeq{f2} where $\epsilon,\gamma$, and $\epsilon'$
are non-zero $\mathcal{O}(1)$ numbers\footnote{We shall not spell out how exactly they acquire such
VEVs. It is however natural to assume that $\mathbf{\bar{6}}$-Higsses with identical $U(1)_\text A$-charges,
living on the same brane, will all acquire the same VEV.}. If the vacuum is supersymmetric,
according to Eq.~(\ref{r7}), the VEVs must be related as \beq \psi~=~\phi\,\sqrt{\tfrac{1}{2}
(\epsilon^2+3\gamma^2+\epsilon'^2)}~.\eeq{f3} On top of this, now we also need to make
sure that the $D$-term corresponding to $U(1)_\text A$ gauge symmetry vanishes. This gives
\beq \langle D_{U(1)_\text A}\rangle~=~\xi_\text{FI}+(+1)\psi^2+\phi^2\left\{3(+1)\gamma^2+(-1)\epsilon'^2\right\}
~=~0,\eeq{f4} where $\xi_\text{FI}$ is the Fayet-Iliopoulos term. In view of~(\ref{f3}), the above
implies \beq \xi_\text{FI}~=~\tfrac{1}{2}\phi^2(\epsilon'^2-\epsilon^2-9\gamma^2).\eeq{f5}
Furthermore, this has to be positive, because $\xi_\text{FI}\propto\sum q=\tfrac{1}{2}>0$.

In the next section we will see that the following choice of VEVs are acceptable in that they,
along with judicial choices of other parameters, can give rise to a realistic pattern for
fermion masses and mixings. \beq \epsilon=1~,~~\gamma=\tfrac{1}{2}~,~~\epsilon'=\tfrac{5}{2}~;
\qquad \phi~\sim~\tfrac{1}{5}M_*~,\qquad \kappa~\sim~\tfrac{1}{3}~.\eeq{f6} Then $\psi$ is
determined from~(\ref{f3}), and $\sigma$ from~(\ref{r13}), so that we have \beq \phi_i~\sim~
\tfrac{1}{5}M_*\left(0,\,0,\,0,\,0,\,0,\,1,\,\tfrac{1}{2},\,\tfrac{1}{2},\,\tfrac{1}{2},\,
\tfrac{5}{2}\right)~;\qquad \psi~\sim~\tfrac{2}{5}M_*~;\qquad \sigma~\sim~\tfrac{4}{45}M_*~.\eeq{f7}
Also Eq.~(\ref{f5}) gives that the Fayet-Iliopoulos term is positive: \beq \xi_\text{FI}~\sim~
\tfrac{3}{2}\phi^2~\sim~\tfrac{3}{50}M_*^2~>~0~.\eeq{f8} From~(\ref{f7}) we find that the GUT
scale is lower than the 5D Planck mass by one order of magnitude: $M_\text{G}\sim\tfrac{1}{10}M_*$.
We also have that $\sqrt{\xi_\text{FI}}\gtrsim M_\text{G}$\,.

To see this is compatible with our model, where we have one extra dimension, we note that the
inverse compactification radius is given by~\cite{add}: \beq r^{-1}~=~2\pi(M_*^3/M_\text P^2),\eeq{f9}
which is of the order of the GUT scale itself: $r^{-1}\sim M_\text{G}$. Therefore, the separation
$d$ between the two branes could still be taken to be larger than $M_*^{-1}$.

The breaking scale of $SU(6)$ is set by the maximum among all the VEVs $\phi_i$ and $\psi$;
it is: $\phi_{10}=\epsilon'\phi\sim\tfrac{1}{2}M_*$. Therefore, as it should be, breaking
of $SU(6)$ takes place below $M_*$, but above $M_\text{G}$. This is in accordance with the choice
of $\kappa\sim\tfrac{1}{3}$, quite expectedly. The light Higgses will then be predominantly
contained in $\Sigma$.

In passing to the next section we parenthetically comment that it is fair to suspect the validity
of field theory description at scales so close to the fundamental Planck mass. Here, however, we
adopt the philosophy of~\cite{Cheng}, i.e. lacking any knowledge whatsoever of how to describe
the full quantum gravity theory, one can assume that even at scales just below $M_*$ the usual
field theory description is valid, and that the gravitational effects can still be incorporated
in Planck-suppressed higher dimensional operators.

\section*{4. Fermion Masses and Mixing Angles}

In this section\footnote{This section goes somewhat parallel to the similar one appearing in~\cite{Cheng}.},
we will work in the units of $M_*=1$. As we know, the light fermion masses arise from non-renormalizable
operators, which are suppressed by powers of $M_*$. After integrating out the extra dimension and the heavy
vector-like fields, one will be left at low energy with an effective 4D Lagrangian $-$ the minimal supersymmetric
standard model (MSSM). The resulting Yukawa couplings among the light matter fields and the light Higgses
will generate fermion masses when the pair of Higgs doublets acquire VEVs. One of the VEVs gives mass to the
up-type quarks, and the other to the down-type quarks and the charged leptons.

The Yukawa couplings will contain various suppression factors. First, (except for the top quark) they
will be suppressed by a VEV $-$ $\phi_i,
\psi$, or $\sigma$ $-$ whenever the corresponding Higgs does \emph{not} father the light Higgses. Second,
if the light Higgs doublets do \emph{not} originate from $\Sigma$, there will be a suppression by $\kappa$.
Third, if fields from \emph{both} the branes are involved, we need to integrate out heavy vector-like fields
(and KK-excitations thereof), which will generate additional suppression. Assuming that the masses of the
(zero modes of the) vector-like fields are smaller than the inverse inter-brane separation ($1/d$), we get
power suppression in $d$~\cite{Cheng}. Let us denote the resulting (dimensionless) suppression factors as
$\beta_6, \beta_{70}$, and $\beta_{20}$, where the subscripts refer to the $SU(6)$ representation. Some
of these can be $\mathcal O(1)$ if we have just one extra dimension~\cite{Cheng}. Finally, whenever an
\emph{external} bulk field is involved, there is a suppression by the extra dimensional volume factor.
For one extra dimension, it is given by~\cite{Cheng, add}: $\theta=(M_*r)^{-1/2}$. With the choice of
VEVs of the previous section, we have that $\theta\sim\tfrac{1}{3}$~.

Because of the charge assignment~(\ref{f1}), among all the $\bar{\Phi}_i$'s only $\bar{\Phi}_6$ and
$\bar{\Phi}_{10}$ may appear below. Let us denote them as $\bar{\Phi}$ and $\bar{\Phi}'$ respectively.
Our conventions for the diagrams are that fields on brane-1 appear on the left, while those on brane-2
appear on the right. The bulk matter fields come with external vertical lines, and the heavy vector-like
fields appear through internal lines between interaction points.
\subsubsection*{4.1. Decoupling of the heavy states}
Using the $SU(3)$ global rotation among the $\zeta _i$'s, we can define $\zeta _3$ as
the one appearing in Fig 1.1, and hence in the operator $\xi\bar{\Phi}'\Psi\zeta _3$.
Similarly, $\mathbf{20_1}$ can be the one which couples to $\xi$ and $\bar{\Phi}'$.
In terms of $SU(5)$ representations: $\mathbf{20}(\xi)=\mathbf{10}+\mathbf{\overline{10}}$,
and $\mathbf{15}(\zeta_i)=\mathbf{10}+\mathbf{5}$. Thus the $\mathbf{{\overline{10}}}$
of $\xi$, and the $\mathbf{10}$ of $\zeta_{3}$ acquire a heavy mass of $\mathcal{O}
(\theta\beta_{20}\epsilon'\phi\psi)$ from this operator\footnote{The various suppression
factors may bring the mass slightly lower than $M_\text{G}$. The gauge coupling unification
however does not get affected as we are considering complete $SU(5)$ multiplets.}.
\vspace{-.5cm}
\begin{center}
\begin{picture}(1000,100) (100,170)
\Text(155,250)[]{$\xi$}\Line(160,190)(190,220)
\Text(210,169)[]{Figure 1.1}\Line(190,220)(160,250)\Line(190,220)(220,220)\Text(205,228)[]{$20_1$}
\Text(220,220)[]{$\times$}\Text(235,228)[]{$20^{\prime}_1$}\Line(220,220)(250,220)\Vertex(250,220){1.5}
\Line(250,220)(250,190)\Text(250,185)[]{$\zeta_3$}\Line(250,220)(280,220)\Text(285,220)[]{$\Psi$}
\Text(155,190)[]{$\bar{\Phi}^{\prime}$}
\Vertex(190,220){1.5}
\end{picture}
\begin{picture}(1000,50) (-110,170)
\Text(190,300)[]{$\zeta_i$}\Line(190,246)(190,270)
\Text(190,240)[]{$\bar{\chi}^{\prime}_j$}\Line(190,270)(190,292)
\Text(190,220)[]{Figure 1.2}\Text(227,270)[]{$\bar{\Phi}$}\Line(190,270)(220,270)
\Vertex(190,270){1.5}
\end{picture}
\end{center}
\vspace{-1.5cm}

In considering Fig 1.2, corresponding the operator $\zeta_i\bar{\Phi}\bar{\chi}'_j$,
we note the decomposition: $\mathbf{\bar6}(\bar{\chi}'_j)=\mathbf{\bar5}+\mathbf{1}$.
The operator $\zeta_i\bar{\Phi}\bar{\chi}'_j$ then gives $SU(5)$-invariant masses of
$\mathcal{O}(\theta^2\epsilon\phi)$ to the heavy states ($\mathbf{5}_{\zeta_i},\mathbf
{\bar5}_{\bar{\chi}'_j}$), which therefore also decouple.

We see that the full $\zeta_3$ becomes heavy. The $SU(5)$ non-singlet fields that survive
as light are only the three $\mathbf{10}$'s coming from ($\xi, \zeta_1, \zeta_2$), and the
three $\mathbf{\bar5}$'s contained in ($\bar{\chi}_1, \bar{\chi}_2, \bar{\chi}_3$); they
contain the three light generations of SM fermions. We postpone for later the discussion
of $SU(5)$ singlet fields in ($\bar{\chi}_i, \bar{\chi}'_i$), all of which decouple as well.

Below we consider the effective operators responsible for the light fermion masses and
mixings. We shall be writing the operators in the $SU(6)$ language. However, it should be
understood that they actually mean to represent only the light fields contained therein.

\subsubsection*{4.2. Yukawa couplings of the up-type quarks}
\vspace{-1cm}
\begin{center}
\begin{picture}(1000,50) (100,280)
\Text(155,300)[]{$\xi$}\Line(160,240)(190,270)
\Text(155,240)[]{$\xi$}\Line(190,270)(160,300)
\Text(175,214)[]{Figure 2.1}\Text(155,270)[]{$\Sigma$}\Line(160,270)(190,270)
\Vertex(190,270){1.5}
\end{picture}
\begin{picture}(1000,50) (-60,180)
\Text(155,250)[]{$\bar{\Phi}^{\prime}$}\Line(160,190)(190,220)
\Text(210,165)[]{Figure 2.2}\Text(155,220)[]{$\Sigma$}\Line(190,220)(160,250)
\Line(190,220)(220,220)\Text(205,228)[]{$70_1$}\Text(220,220)[]{$\times$}\Text(235,228)[]
{$\overline{70}_1$}\Line(220,220)(250,220)\Vertex(250,220){1.5}\Line(250,220)(250,190)
\Text(250,185)[]{$\zeta_2$}\Line(250,220)(280,220)\Text(285,220)[]{$\Psi$}
\Text(155,190)[]{$\xi$}\Line(160,220)(190,220)
\Vertex(190,220){1.5}
\end{picture}
\end{center}
\vspace{0.5cm}
The operator $\xi\Sigma\xi$, corresponding to Fig 2.1, gives to the top quark Yukawa
coupling an $\mathcal{O}(1)$ contribution with no suppression factors. In fact,
in the first place, we have chosen $\xi$ to reside on the same brane as $\Sigma$
in order for this to happen.

We have exploited the rotation freedom between ($\zeta_1, \zeta_2$) to define $\zeta _{2}$
as the only one that couples to $\mathbf{\overline{70}_1}$ and $\Psi$, as is seen in Fig
2.2. The corresponding operator $\xi \Sigma \bar{\Phi}^{\prime }\Psi \zeta _{2}$ gives
rise to the 23 and 32 elements of the Yukawa coupling matrix of the up-type quarks.
The resulting contribution is of $\mathcal{O}(\theta\beta_{70}\epsilon'\phi\psi)$.\newline
\newline

\vspace{-1.2cm}
\begin{center}
\begin{picture}(1000,50) (140,240)
\Text(155,250)[]{$\bar{\Phi}^{\prime}$}\Line(160,190)(190,220)
\Text(210,134)[]{Figure 2.3}\Text(155,220)[]{$\xi$}\Line(190,220)(160,250)\Line(190,220)(220,220)
\Text(205,228)[]{$20_2$}\Text(220,220)[]{$\times$}\Text(235,228)[]{$20^{\prime}_2$}
\Line(220,220)(250,220)\Vertex(250,220){1.5}\Line(250,220)(250,190)\Text(250,185)[]
{$\zeta_i$}\Line(250,220)(280,220)\Text(285,220)[]{$\bar{\Phi}$}
\Text(155,190)[]{$\bar{\Phi}^{\prime}$}\Line(160,220)(190,220)
\Line(250,220)(280,250)\Text(285,255)[]{$\Psi$}
\Line(250,220)(280,190)\Text(287,187)[]{$\Psi$}
\Vertex(190,220){1.5}
\end{picture}
\vspace{3cm}
\begin{picture}(1000,50) (-60,240)
\Text(155,300)[]{$\bar{\Phi}^{\prime}$}\Line(160,240)(190,270)
\Text(155,240)[]{$\bar{\Phi}^{\prime}$}\Line(190,270)(160,300)
\Text(210,185)[]{Figure 2.4}\Text(155,270)[]{$\Sigma$}\Line(160,270)(190,270)
\Vertex(190,270){1.5}
\Line(190,270)(250,304.641)\Text(200,286)[]{$70_1$}\Text(220,288)[]{$\times$}
\Vertex(250,304.641){1.5}\Text(230,304)[]{$\overline{70}_1$}\Line(190,270)(250,235.359)
\Text(200,255)[]{$20_1$}\Text(220,254)[]{$\times$}\Text(230,237)[]{$20^{\prime}_1$}
\Vertex(250,235.359){1.5}\Line(250,304.641)(250,334.641)\Text(250,342)[]{$\zeta_2$}
\Line(250,304.641)(280,304.641)\Text(285,306)[]{$\Psi$}\Line(250,235.359)(250,205.359)
\Text(250,200)[]{$\zeta_i$}\Line(250,235.359)(280,235.359)\Text(285,237.359)[]{$\Psi$}
\end{picture}
\end{center}
\vspace{2cm}

The operator $\xi \bar{\Phi}^{\prime }\bar{\Phi}^{\prime }\bar{\Phi}\Psi \Psi \zeta _{i}$,
corresponding to Fig 2.3, will provide contributions of  $\mathcal{O}(\kappa\theta\beta_{20}
\epsilon\epsilon'^2\phi^3\psi)$ to the 13, 31, 23 and 32 elements of the Yukawa coupling
matrix. As the light Higgs did not come from $\Sigma$, there has been a suppression by
the mixing angle $\kappa$.

What appears in Fig 2.4 is the operator $\zeta _2\Psi\bar\Phi'\Sigma\bar\Phi'\Psi\zeta _i$,
which generates the 12, 21 and 22 elements of the Yukawa coupling matrix of $\mathcal{O}
(\theta^2\beta_{20}\beta_{70}\epsilon'^2\phi^2\psi^2)$.

Finally, the leading contribution of $\mathcal{O}(\theta^2\beta_{20}\beta_{70}\epsilon\epsilon'^3
\phi^4\psi^3)$ to the 11 element of the up-type Yukawa matrix come from the operator $\zeta_i
\bar{\Phi}\Psi\Sigma\bar{\Phi}'\bar{\Phi}'\bar{\Phi}'\Psi\Psi\zeta_j$ of Fig 2.5. It is the most
suppressed among all the up-type Yukawa couplings.

To summarize, at leading order the various elements of the up-type quark Yukawa coupling matrix
at the GUT scale are given by: \bea &[\lambda_U]_{11}~\sim~\mathcal{O}(\theta^2\beta_{20}\beta_{70}\epsilon
\epsilon'^3\phi^4\psi^3)~,&\label{y1}\\&[\lambda_U]_{12}~,~[\lambda_U]_{21}~,~[\lambda_U]_{22}~\sim~\mathcal{O}(\theta^2
\beta_{20}\beta_{70}\epsilon'^2\phi^2\psi^2)~,&\label{y2}\\&[\lambda_U]_{13}~,~[\lambda_U]_{31}~\sim~\mathcal{O}(\kappa
\theta\beta_{20}\epsilon\epsilon'^2\phi^3\psi)~,&\label{y3}\\&[\lambda_U]_{23}~,~[\lambda_U]_{32}~\sim~\mathcal{O}(\theta
\beta_{70}\epsilon'\phi\psi)+\mathcal{O}(\kappa\theta\beta_{20}\epsilon\epsilon'^2\phi^3\psi)~,&\label{y4}\\
&[\lambda_U]_{33}~\sim~\mathcal{O}(1)~.&\eea{y5}

\begin{picture}(1000,50) (30,350)
\Text(155,230)[]{$\bar{\Phi}^{\prime}$}\Line(190,270)(160,230)
\Text(155,250)[]{$\bar{\Phi}^{\prime}$}\Line(190,270)(160,252)
\Text(155,295)[]{$\bar{\Phi}^{\prime}$}\Line(190,270)(160,295)
\Text(210,190)[]{Figure 2.5}\Text(155,270)[]{$\Sigma$}\Line(160,270)(190,270)
\Vertex(190,270){1.5}
\Line(190,270)(250,304.641)\Text(200,286)[]{$70_1$}\Text(220,288)[]{$\times$}
\Vertex(250,304.641){1.5}\Text(230,304)[]{$\overline{70}_1$}\Line(190,270)(250,235.359)
\Text(200,255)[]{$20_2$}\Text(220,254)[]{$\times$}\Text(230,237)[]{$20^{\prime}_2$}
\Vertex(250,235.359){1.5}\Line(250,304.641)(250,334.641)\Text(250,342)[]{$\zeta_i$}
\Line(250,304.641)(280,304.641)\Text(285,306)[]{$\Psi$}\Line(250,304.641)(280,334.641)
\Text(284,341)[]{$\bar{\Phi}$}\Line(250,235.359)(250,205.359)\Text(250,200)[]{$\zeta_j$}
\Line(250,235.359)(280,235.359)\Text(285,237.359)[]{$\Psi$}
\Line(250,235.359)(280,205.359)\Text(285,203)[]{$\Psi$}
\end{picture}
\vspace{6cm}

\subsubsection*{4.3. Yukawa couplings of the down-type quarks}

\vspace{1cm}
\begin{center}
\begin{picture}(1000,50) (130,220)
\Text(155,250)[]{$\Sigma$}\Line(160,190)(190,220)
\Text(210,134)[]{Figure 3.1}\Line(190,220)(160,250)\Line(190,220)(220,220)
\Text(205,228)[]{$70_0$}\Text(220,220)[]{$\times$}\Text(235,228)[]{$\overline{70}_0$}
\Line(220,220)(250,220)\Vertex(250,220){1.5}\Line(250,220)(250,190)\Text(250,185)[]
{$\bar{\chi}_3$}\Line(250,220)(280,250)\Text(285,255)[]{$\bar{\Phi}$}
\Line(250,220)(280,190)\Text(287,187)[]{$\bar{\Phi}$}
\Text(155,190)[]{$\xi$}
\Vertex(190,220){1.5}
\end{picture}\vspace{2cm}
\begin{picture}(1000,50) (-60,220)
\Text(155,300)[]{$\bar{\Phi}^{\prime}$}\Line(160,240)(190,270)
\Text(155,240)[]{$\Sigma$}\Line(190,270)(160,300)
\Vertex(190,270){1.5}
\Line(190,270)(250,304.641)\Text(200,286)[]{$70_1$}\Text(220,288)[]{$\times$}
\Vertex(250,304.641){1.5}\Text(230,304)[]{$\overline{70}_1$}
\Line(190,270)(250,235.359)\Text(200,255)[]{$\overline{70}_0$}\Text(220,254)[]{$\times$}
\Text(230,238)[]{$70_0$}\Vertex(250,235.359){1.5}
\Line(250,304.641)(250,334.641)\Text(250,342)[]{$\zeta_2$}\Line(250,304.641)(280,304.641)
\Text(285,306)[]{$\Psi$}
\Line(250,235.359)(250,205.359)\Text(253,200)[]{$\bar{\chi}_3$}\Line(250,235.359)(280,235.359)
\Text(285,237.359)[]{$\bar{\Phi}$}
\Line(250,235.359)(280,205.359)\Text(285,203)[]{$\bar{\Phi}$}
\Text(210,185)[]{Figure 3.2}
\end{picture}
\end{center}
\vspace{1.5cm}

In Fig 3.1 we have used the $SU(3)$ global rotation among the $\bar{\chi}_i$'s, so that only $\bar{\chi}_3$
appears therein, and therefore in the operator $\xi\Sigma\bar{\Phi}\bar{\Phi}\bar{\chi}_3$. The latter
gives a leading contribution of $\mathcal{O}(\theta\beta_{70}\epsilon^2\phi^2)$ to the 33 element of the
down-type quark Yukawa matrix\footnote{It also generates the same of the charged leptons, which we will
discuss in the next subsection.}.

The operator $\bar{\chi}_3\bar{\Phi}\bar{\Phi}\Sigma\bar{\Phi}'\Psi\zeta _2$, corresponding to Fig 3.2,
will generate the 23 element of the Yukawa coupling matrix. The contribution is of $\mathcal{O}(\theta^2
\beta_{70}^2\epsilon^2\epsilon'\phi^3\psi)$.

\vspace{-1cm}
\begin{center}
\begin{picture}(1000,30) (150,250)
\Text(220,160)[]{Figure 3.3}\Text(155,220)[]{$\Sigma$}\Line(190,220)(220,220)
\Text(205,228)[]{$6_0$}\Text(220,220)[]{$\times$}\Text(235,228)[]{$\overline{6}_0$}
\Line(220,220)(250,220)\Vertex(250,220){1.5}\Line(250,220)(250,190)\Text(252,184)[]{$\zeta_j$}
\Line(250,220)(280,220)\Text(286,221)[]{$\bar{\Phi}$}
\Text(192,185)[]{$\bar{\chi}_i$}\Line(160,220)(190,220)
\Line(190,220)(190,190)
\Vertex(190,220){1.5}
\end{picture}
\begin{picture}(1000,49) (-60,200)
\Text(155,250)[]{$\bar{\Phi}^{\prime}$}\Line(160,190)(190,220)
\Text(220,160)[]{Figure 3.4}\Line(190,220)(160,250)\Line(190,220)(220,220)\Text(205,228)[]{$6_1$}
\Text(220,220)[]{$\times$}\Text(235,228)[]{$\overline{6}_1$}\Line(220,220)(250,220)
\Vertex(250,220){1.5}\Line(250,220)(250,190)\Text(252,184)[]{$\zeta_j$}\Line(250,220)(280,220)
\Text(285,220)[]{$\Psi$}
\Line(190,220)(190,190)\Text(195,184)[]{$\bar{\chi}_{2,3}$}
\Line(250,220)(280,250)\Text(285,255)[]{$\bar{\Phi}$}
\Line(250,220)(280,190)\Text(287,187)[]{$\bar{\Phi}$}
\Text(155,190)[]{$\Sigma$}
\Vertex(190,220){1.5}
\end{picture}
\end{center}
\vspace{1.7cm}

The operator in Fig 3.3 has the intermediate states $(\mathbf6, \mathbf{\bar6})$, which do
\emph{not} contain the $(\mathbf{10}, \mathbf{\overline{10}})$ of $SU(5)$. For any such operator,
the light Higgses \emph{cannot} come from $\Sigma$. The operator $\langle\Sigma\rangle\zeta _j
\bar{\Phi}\bar{\chi}_i$ is however irrelevant for the light fermion masses, in view of the
operator $\zeta _i\bar{\Phi}\bar{\chi}_j$ in Fig 1.2; the former only redefines the composition
of the heavy fermions.

On the other hand, the operator $\langle\Sigma\rangle\bar{\chi}_{2,3}\bar{\Phi}'\bar{\Phi}
\bar{\Phi}\Psi\zeta _j$ in Fig 3.4, \emph{does} contribute to the light fermion masses.
Note that we have used the rotation freedom of $\bar{\chi}_i$'s, so that only $\bar{\chi}_{2,3}$
appear in the above. Because we have already exhausted in the up sector the rotation freedom
between $(\zeta_1, \zeta_2)$, this operator contributes to all the 12, 13, 22 and 23 elements
of the down-type Yukawa matrix. The contribution is of $\mathcal{O}(\kappa\theta^2\beta_6\epsilon^2
\epsilon'\phi^3\sigma)$. However, compared to the (22, 23) elements, the elements (12, 13) are
further suppressed by the Cabibbo-mixing term: $\sin\theta_{\text{C}}\sim0.22$\,.

In what follows we consider the leading contributions to the other elements of the down-type
Yukawa matrix. They are more suppressed than the above diagrams.

\vspace{-2.7cm}
\begin{center}
\begin{picture}(1000,50) (40,285)
\Text(155,220)[]{$\Sigma$}\Line(160,220)(190,220)
\Text(155,250)[]{$\bar{\Phi}^{\prime}$}\Line(160,190)(190,220)
\Text(220,160)[]{Figure 3.5}\Line(190,220)(160,250)\Line(190,220)(220,220)\Text(205,228)[]
{$\overline{70}_1$}\Text(220,220)[]{$\times$}\Text(235,228)[]{$70_1$}\Line(220,220)(250,220)
\Vertex(250,220){1.5}\Line(250,220)(280,228.03848)\Text(285,228.03848)[]{$\bar{\Phi}$}
\Line(250,220)(280,211.962)\Text(285,211.962)[]{$\bar{\Phi}$}
\Line(250,220)(280,250)\Text(285,255)[]{$\Psi$}
\Line(250,220)(280,190)\Text(287,187)[]{$\bar{\Phi}$}
\Text(155,190)[]{$\xi$}
\Line(250,220)(250,190)\Text(252,185)[]{$\bar{\chi}_i$}
\Vertex(190,220){1.5}
\end{picture}
\end{center}
\vspace{4.5cm}

The operator corresponding to Fig 3.5 is $\xi\Sigma\bar{\Phi}'\bar{\Phi}\bar{\Phi}\bar{\Phi}
\Psi\bar{\chi}_{i}$, which gives the leading contributions of $\mathcal{O}(\theta\beta_{70}
\epsilon^3\epsilon'\phi^4\psi)$ to the 31 and 32 elements of the down-type Yukawa matrix.

In Fig 3.6, we have the operator $\bar\chi_i\bar\Phi\bar\Phi\bar\Phi\bar\Psi\bar\Phi'\Sigma
\bar\Phi'\Psi\zeta_2$; it gives the leading contribution of $\mathcal{O}(\theta^2\beta_{70}^2
\epsilon^3\epsilon'^2\phi^5\psi^2)$ to the 21 element of the down-type quark Yukawa matrix.
\vspace{-3cm}
\begin{center}
\begin{picture}(1000,50) (140,380)
\Text(155,300)[]{$\bar{\Phi}^{\prime}$}\Line(160,240)(190,270)
\Text(155,240)[]{$\bar{\Phi}^{\prime}$}\Line(190,270)(160,300)
\Text(210,189)[]{Figure 3.6}\Text(155,270)[]{$\Sigma$}\Line(160,270)(190,270)
\Vertex(190,270){1.5}
\Line(190,270)(250,304.641)\Text(200,286)[]{$70_1$}\Text(220,288)[]{$\times$}
\Vertex(250,304.641){1.5}\Text(230,304)[]{$\overline{70}_1$}\Line(190,270)(250,235.359)
\Text(200,255)[]{$\overline{70}_1$}\Text(220,254)[]{$\times$}\Text(230,237)[]{$70_1$}
\Vertex(250,235.359){1.5}\Line(250,304.641)(250,334.641)\Text(250,342)[]{$\zeta_2$}
\Line(250,304.641)(280,304.641)\Text(285,306)[]{$\Psi$}\Line(250,235.359)(250,205.359)
\Text(252,200)[]{$\bar{\chi}_i$}\Line(250,235.359)(280,235.359)\Text(285,237.359)[]{$\Psi$}
\Line(250,235.359)(280,222.933)\Text(285,223)[]{$\bar{\Phi}$}
\Line(250,235.359)(280,205.359)\Text(285,205)[]{$\bar{\Phi}$}
\Line(250,235.359)(262.426,205.359)\Text(266,200)[]{$\bar{\Phi}$}
\end{picture}
\begin{picture}(1000,50) (-60,280)
\Text(155,220)[]{$\bar{\Phi}^{\prime}$}\Line(160,220)(190,220)
\Text(155,250)[]{$\Sigma$}\Line(160,190)(190,220)
\Text(220,140)[]{Figure 3.7}\Line(190,220)(160,250)\Line(190,220)(220,220)\Text(205,228)[]
{$6_1$}\Text(220,220)[]{$\times$}\Text(235,228)[]{$\overline{6}_1$}\Line(220,220)(250,220)
\Vertex(250,220){1.5}\Line(250,220)(250,190)\Text(252,184)[]{$\zeta_j$}\Line(250,220)(280,220)
\Text(285,220)[]{$\Psi$}
\Line(190,220)(190,190)\Text(192,184)[]{$\bar{\chi}_{i}$}
\Line(250,220)(280,250)\Text(285,255)[]{$\bar{\Phi}$}
\Line(250,220)(280,190)\Text(287,187)[]{$\bar{\Phi}$}
\Text(155,190)[]{$\Sigma$}
\Vertex(190,220){1.5}
\end{picture}
\end{center}
\vspace{5.2cm}

Finally, the operator $\langle\Sigma\rangle^2\bar\chi_i\bar\Phi'\bar\Phi\Psi\bar\Phi\zeta_j$,
corresponding to Fig 3.7 generates the leading contribution of $\mathcal{O}(\kappa\theta^2
\beta_6\sigma^2\epsilon^2\epsilon'\phi^3)$ to the 11 element of the Yukawa matrix .

Summarizing, at leading order the various elements of the down-type Yukawa coupling matrix at
the GUT scale are given by:
\begin{eqnarray}
&[\lambda_D]_{11}~\sim~\mathcal{O}(\kappa\theta^2\beta_6\sigma^2\epsilon^2\epsilon'\phi^3)~,&\label{yd1}\\
&[\lambda_D]_{12}~,~[\lambda _{D}]_{13}~\sim~\mathcal{O}(\kappa\theta^2\beta_6\epsilon^2\epsilon'\phi^3\sigma
\sin\theta_{\text{C}})~,&\label{yd2}\\&[\lambda_D]_{21}~\sim~\mathcal{O}(\theta^2\beta_{70}^2\epsilon^3\epsilon'^2\phi^5
\psi^2)~,&\label{yd3}
\end{eqnarray}
\begin{eqnarray}
&[\lambda _{D}]_{22}~\sim~\mathcal{O}(\kappa\theta^2\beta_6\epsilon^2\epsilon'\phi^3\sigma)~,& \label{yd4}\\
&[\lambda _{D}]_{23}~\sim~\mathcal{O}(\theta^2\beta_{70}^2\epsilon^2\epsilon'\phi^3\psi)+\mathcal{O}(\kappa\theta^2\beta_6
\epsilon^2\epsilon'\phi^3\sigma)~,&\label{yd5}\\&[\lambda _{D}]_{31}~,~[\lambda _{D}]_{32}~\sim~\mathcal{O}(\theta
\beta_{70}\epsilon^3\epsilon'\phi^4\psi)~,&\label{yd6}\\&[\lambda _{D}]_{33}~\sim~\mathcal{O}(\theta\beta_{70}
\epsilon^2\phi^2)~.&\label{yd7}
\end{eqnarray}

\subsubsection*{4.4. Yukawa couplings of the charged leptons}

The diagrams that generate the Yukawa couplings of the down-type quarks also produce the
Yukawa couplings of the charged leptons. The operator in Fig 3.1, for example, gives similar
contribution to the 33 element of the charged lepton Yukawa matrix. This accounts for why one
may have approximate $b-\tau$ unification at the GUT scale.

From the VEV $\langle\Sigma_\alpha^\beta\rangle = {\text{diag}}(1,1,1,1,-2,-2)\,\sigma$,
it is clear that whenever a $\langle\Sigma\rangle$ appears in an operator, the charged lepton Yukawa
matrix elements will have a factor $-2$ compared their down-type quark counterparts. Now the leading
contribution to the second generation masses come from one such operator, namely $\langle\Sigma
\rangle\bar{\chi}_{2,3}\bar{\Phi}'\bar{\Phi}\bar{\Phi}\Psi\zeta _j$, i.e. that of  Fig 3.4.
This provides an explanation to the $s-\mu$ mass discrepancy at the GUT scale.

The various elements of the charged lepton Yukawa coupling matrix, at leading order, at the GUT
scale are given by:
\begin{eqnarray}
&[\lambda_l]_{11}~\sim~\mathcal{O}(4\kappa\theta^2\beta_6\sigma^2\epsilon^2\epsilon'\phi^3)~,&\label{yl1}\\
&[\lambda_l]_{12}~,~[\lambda_{l}]_{13}~\sim~-\mathcal{O}(2\kappa\theta^2\beta_6\epsilon^2\epsilon'\phi^3
\sigma)~,&\label{yl2}\\&[\lambda_l]_{21}~\sim~\mathcal{O}(\theta^2\beta_{70}^2\epsilon^3\epsilon'^2\phi^5
\psi^2)~,&\label{yl3}\\&[\lambda_l]_{22}~\sim~-\mathcal{O}(2\kappa\theta^2\beta_6\epsilon^2\epsilon'\phi^3
\sigma)~,& \label{yl4}\\&[\lambda_l]_{23}~\sim~\mathcal{O}(\theta^2\beta_{70}^2\epsilon^2\epsilon'\phi^3\psi)
-\mathcal{O}(2\kappa\theta^2\beta_6\epsilon^2\epsilon'\phi^3\sigma)~,&\label{yl5}\\&[\lambda_l]_{31}~,~[\lambda_l]_{32}
~\sim~\mathcal{O}(\theta\beta_{70}\epsilon^3\epsilon'\phi^4\psi)~,&\label{yl6}\\&[\lambda_l]_{33}~\sim~
\mathcal{O}(\theta\beta _{70}\epsilon^2\phi^2)~.&\label{yl8}
\end{eqnarray}

\subsubsection*{4.5. Fermion masses and CKM matrix}

To find the fermion masses we diagonalize the matrices $\lambda_U, \lambda_D$, and $\lambda_l$.
We denote the corresponding eigenvalues as follows: $\lambda_U\rightarrow\text{diag}(\lambda_u,
\lambda_c,\lambda_t)$, $\lambda_D\rightarrow\text{diag}(\lambda_d,\lambda_s,\lambda_b)$, and
$\lambda_l\rightarrow\text{diag}(\lambda_e,\lambda_\mu,\lambda_\tau)$. The eigenvalues are
nothing but the GUT-scale Yukawa couplings for the mass eigenstates of the up-type quarks, the
down-type quarks, and the charged leptons respectively. They are related to the corresponding masses in the
following way. \bea m_i&=&\tfrac{1}{\sqrt2}\,v\sin\beta\lambda_i\qquad i=u,c,t~,\label{m1}\\
m_j&=&\tfrac{1}{\sqrt2}\,v\cos\beta\lambda_j\qquad j=d,s,b,e,\mu,\tau~,\eea{m2} where $\tan\beta$
is the ratio of the Higgs-doublet VEVs, which respectively give mass to the up-type quarks and to
the down-type quarks (and the charged leptons), and $v\approx246\,\text{GeV}$.

One can choose the various parameters in the model to evaluate the fermion masses at the GUT scale,
and then compare them with some standard reference values. With the choice of VEVs given
in~(\ref{f6}-\ref{f7}), we further choose the remaining parameters as: \beq \theta~\sim~\tfrac{1}{3}~,
\qquad \beta_6~\sim~1~,\qquad \beta_{70}~\sim~\tfrac{1}{2}~,\qquad \beta_{20}~\sim~\tfrac{1}{40}~.\eeq{m3}
In Table~\ref{tab:table1}, we give the GUT-scale fermion masses, evaluated with the above choice of
parameters\footnote{We already argued that these choices are compatible with our model that has
one extra dimension.}, and with $\tan\beta=2$\,. We also provide with the GUT-scale reference values,
obtained by renormalization group running in the context of MSSM, for $\tan\beta=2$~\cite{tan1}.

\begin{table}[ht!]
\begin{center}
\begin{tabular}{||c|c|c||}
\hline\hline
GUT-Scale  & Predicted Value & Reference Value \\
Mass & ($\tan\beta=2$) & ($\tan\beta=2$) \\ \hline\hline
$m_u$ (MeV) & $\sim 0.8$ & $0.679^{+0.308}_{-0.277}$  \\
$m_c$ (GeV) & $\sim 0.2$ & $0.309^{+0.046}_{-0.056}$ \\
$m_t$ (GeV) & $\sim 150$ & $148^{+38}_{-23}$ \\ \hline
$m_d$ (MeV) & $\sim 0.5$ & $0.731^{+0.221}_{-0.209}$ \\
$m_s$ (MeV) & $\sim 4$ & $16.0^{+4.6}_{-5.6}$ \\
$m_b$ (GeV) & $\sim 0.5$ & $0.929^{+0.073}_{-0.044}$ \\ \hline
$m_e$ (MeV) & $\mathcal{O}(1)$ & 0.312 \\
$m_{\mu}$ (MeV) & $\sim 10$ & 65.9 \\
$m_{\tau}$ (GeV) & $\sim 0.5$ & 1.12 \\ \hline\hline
\end{tabular}%
\caption{Predicted and reference values of fermion masses at the GUT scale.}
\label{tab:table1}
\end{center}
\end{table}
We see that our model mimics remarkably the pattern for the fermion masses at the GUT
scale\footnote{Masses of the first generation are challenging to obtain~\cite{bdsbh},
because they may be affected by many different factors. Our electron mass, in particular,
could have been reported only up to an order of magnitude; there appeared a few contributions
to it of the same order, which might undergo a mild cancelation among themselves to
produce a small electron mass.}. Renormalization group running to low energies will
likewise yield a realistic pattern. The approximation is very good for small $\tan\beta$.
However, one finds that for large $\tan\beta$, say $\tan\beta=10$, the reference
values~\cite{tan1, tan2} do not fit easily into our model.

To be more enthusiastic about the model, let us estimate the Cabibbo-Kobayashi-Maskawa
(CKM) matrix elements. In the basis where the up-type quark Yukawa coupling matrix is
diagonal, the CKM matrix defines the unitary transformation, which when acts on the
down-type quark mass eigenstates, gives the weak eigenstates~\cite{CKM}. Let $O_U$ and
$O_D$ be the matrices, which respectively diagonalize the Yukawa coupling matrices
$\lambda_U$ and $\lambda_D$. Then the CKM matrix $K$ is given by: \beq K~\equiv~O_U^T
P_{UD}O_D~,\qquad P_{UD}~=~\text{diag}\left(1, e^{-i\varphi}, e^{-i\tau}\right),\eeq{ckm1}
where $\varphi, \tau$ are arbitrary phase parameters, which are at our disposal.

Starting from the Yukawa coupling matrices $-$ $\lambda_U$, given in Eqs.~(\ref{y1}-\ref{y5}),
and $\lambda_D$, given in Eqs.~(\ref{yd1}-\ref{yd7}) $-$ one can compute the matrices
$O_U$ and $O_D$ for the parameter values spelled out in~(\ref{f7},\ref{m3}). The CKM
matrix elements at the GUT scale will then be given by Eq.~(\ref{ckm1}), as functions of the
parameters $\varphi$ and $\tau$. We can run the matrix elements down to the $m_Z$ scale by
the standard renormalization group equations~\cite{evol}, and then choose judiciously
$\varphi$ and $\tau$ in order to fit with experimental results.

Apart from the magnitudes of the CKM elements, we are particularly interested in the
CP-violating phase $\delta$, and the Jarlskog invariant $J$, which are defined
as~\cite{Branco,Jarlskog}: \beq \sin\delta~\equiv~\frac{(1-|K_{ub}|^2)\,J}{|K_{ud}K_{us}
K_{ub}K_{cb}K_{tb}|}~,\qquad J~\equiv~\Im(K_{us}K_{cb}K_{ub}^*K_{cs}^*)~,\eeq{ckm2}
where we have used the standard notation:
\beq K~\equiv~\left(
\begin{array}{ccc}
K_{ud} & K_{us} & K_{ub} \\
K_{cd} & K_{cs} & K_{cb} \\
K_{td} & K_{ts} & K_{tb} \\
\end{array}
\right)
\eeq{ckm3}

The various quantities at $m_Z$ predicted from our model have the best agreement with
the experimental values~\cite{PDG} if we adjust the parameters $\varphi$ and $\tau $ as:
\beq \varphi\sim 140^\circ,\qquad \tau\sim 180^\circ.\eeq{ckm4} Table~\ref{tab:table2}
gives the predicted and (central) experimental values~\cite{PDG} at $m_Z$ of the magnitudes
of the CKM elements, the CP-violating phase $\delta$, and the Jarlskog invariant $J$.
\begin{table}[ht!]
\begin{center}
\begin{tabular}{||c|c|c||}
\hline\hline
Quantity & Predicted  & Central \\
at $m_Z$ & Value & Experimental Value\\\hline\hline
$|K_{ud}|$ & $\sim 0.98$ & 0.97419 \\
$|K_{us}|$ & $\sim 0.18$ & 0.2257 \\
$|K_{ub}|$ & $\sim 0.0035$ & 0.00359 \\ \hline
$|K_{cd}|$ & $\sim 0.18$ & 0.2256 \\
$|K_{cs}|$ & $\sim 0.98$ & 0.97334 \\
$|K_{cb}|$ & $\sim 0.053$ & 0.0415 \\ \hline
$|K_{td}|$ & $\sim 0.0073$ & 0.00874 \\
$|K_{ts}|$ & $\sim 0.053$ & 0.0407 \\
$|K_{tb}|$ & $\sim 0.99$ & 0.999133\\ \hline
$\delta$ & $\sim 73^\circ$ & $77^\circ$ \\ \hline
$J$ & $\sim 0.000031$ & 0.0000305 \\ \hline\hline
\end{tabular}%
\caption{Predicted and experimental values of the magnitudes of the CKM elements,
the CP-violating phase $\delta$, and the Jarlskog invariant $J$ at the $m_Z$ scale.}
\label{tab:table2}
\end{center}
\end{table}

Clearly the texture of the quark Yukawa couplings in our model matches quite
nicely with the SM one. Indeed here the agreement of experimental data with our
model is as good as in the models of \cite{Branco2010,Xing2010,Branco2012,Ivo2012,CarcamoHernandez:2012xy,King2013,Hernandez:2013hea,Hernandez:2014vta,Gomez-Izquierdo:2013uaa,Canales:2013cga,Campos:2014lla,Hernandez:2014zsa,Hernandez:2014hka,Hernandez:2015cra,Arbelaez:2015toa,Hernandez:2015dga,CarcamoHernandez2015} and much better than with many others obtained, for example, from various
mass matrix ansatze~\cite{Fritzsch,Xing,FX,Matsuda,Zhou,Carcamo}.

It is remarkable that the feat of obtaining a realistic pattern for the SM fermion
masses and mixings can be achieved at all, without appealing to any flavor symmetry,
just by choosing various mass scales in our model such that appropriate suppression
factors show up naturally. We will finish this section with a short subsection devoted
to the neutrino masses and mixings, for which we will need some additional ingredient,
as we will see.
\subsubsection*{4.6. Majorana masses and mixings of neutrinos}

\vspace{-2cm}
\begin{center}
\begin{picture}(1000,50) (140,280)
\Text(155,250)[]{$\bar{\Phi}'$}\Line(160,190)(190,220)
\Text(220,144)[]{Figure 6.1}\Line(190,220)(160,250)\Line(190,220)(220,220)
\Text(205,227)[]{$20_2$}
\Text(220,220)[]{$\times$}\Text(235,227)[]{$20'_2$}\Line(220,220)(250,220)
\Vertex(250,220){1.5}\Line(250,220)(250,190)\Text(250,185)[]{$\bar{\chi}_j$}
\Line(190,220)(190,190)\Text(195,185)[]{$\bar{\chi}_i$}
\Line(250,220)(280,250)\Text(285,255)[]{$\Psi$}
\Line(250,220)(280,190)\Text(287,187)[]{$\Psi$}
\Text(155,190)[]{$\bar{\Phi}'$}
\Vertex(190,220){1.5}
\end{picture}
\begin{picture}(1000,50) (-60,215)
\Text(220,130)[]{Figure 6.2}
\Line(190,220)(160,250)\Text(155,250)[]{$\bar{\Phi}'$}
\Line(190,190)(160,160)\Text(155,160)[]{$\bar{\Phi}'$}
\Line(250,220)(280,250)\Text(285,255)[]{$\Psi$}
\Line(250,190)(280,160)\Text(285,155)[]{$\Psi$}
\Line(190,220)(220,220)\Text(205,227)[]{$70_1$}\Text(220,220)[]{$\times$}
\Text(235,227)[]{$\overline{70}_1$}
\Line(190,190)(220,190)\Text(205,197)[]{$70_1$}\Text(220,190)[]{$\times$}
\Text(235,197)[]{$\overline{70}_1$}
\Line(220,220)(250,220)
\Line(220,190)(250,190)
\Line(250,190)(250,220)
\Text(250,205)[]{$\times$}
\Text(260,205)[]{$\eta$}
\Vertex(250,220){1.5}
\Vertex(250,190){1.5}
\Vertex(190,220){1.5}
\Vertex(190,190){1.5}
\Line(190,250)(190,220)
\Line(190,160)(190,190)
\Text(190,153)[]{$\bar{\chi}_j$}
\Text(190,257)[]{$\bar{\chi}_i$}
\end{picture}
\end{center}
\vspace{3cm}

The operator corresponding to Fig 6.1 not only provides with Majorana masses to the left
handed neutrinos, but also renders heavy the $SU(5)$ singlets contained in $\bar\chi_i$'s
with masses of $\mathcal O(\theta^2\beta_{20}\epsilon'^2\phi^2\psi^2)$. A similar diagram
gives heavy mass also to the $SU(5)$ singlets in $\bar\chi'_i$'s, so that the $\bar\chi'_i$'s
decouple \emph{completely}. The rotational invariance among the $\bar{\chi}_{i}$'s demands
that there be large neutrino flavor-mixings in general, which is consistent with experimental
data~\cite{neutrino}, except for the third mixing angle: $0<\theta_{13}<13^\circ$. However,
this diagram generates too small a neutrino mass: $m_\nu\sim\kappa^{2}\theta^{2}\beta_{20}
\epsilon'^2\phi^2M_*^{-1}\left(v/\sqrt2\right)^2\sim 10^{-8}\,\text{eV}$, as opposed to the
heaviest neutrino mass $\sim\left(3\times10^{-2}-10^{-1}\right)\,\text{eV}$.

The above problem can be taken care of by introducing on brane-2 a heavy field $\eta$, with
mass $M_\eta$ and matter parity $-1$, which is a singlet of both the gauge groups $SU(6)$ and
$U(1)_{\text A}$. Then the diagram of Fig 6.2 gives a Majorana neutrino mass: \beq m_\nu~\sim~
\kappa^{2}\theta^{2}\beta_{70}^2\epsilon'^2\phi^2M_\eta^{-1}\left(v/\sqrt2\right)^2~,\eeq{neut1}
which can be the heaviest neutrino mass, if $M_\eta\sim10^{12}\,\text{GeV}$ is
chosen\footnote{This scale is 4 orders of magnitude lower than the GUT scale. However, one
usually expects some new physics to kick in \emph{below} the GUT scale in order to account
for the neutrino masses~\cite{neutrino}.}. A lighter neutrino mass can similarly be
obtained by adding another singlet field ($\eta'$) with a larger mass.

\newpage
\subsection*{5. Conclusion}

In this paper we have presented a supersymmetric $SU(6)$ GUT model in which the doublet-triplet
splitting is natural. The model is novel in that the global symmetries of the Higgs superpotential
result from a non-vectorlike Higgs content, and just as such it is worth studying. The explicit
realization of the model involves one flat extra dimension and branes. Localization of the Higgs
fields on separate branes automatically forbids all non-renormalizable terms that could otherwise
ruin the Higgs-sector global symmetry, and as a result a pair of light Higgs doublets appear in
the guise of pseudo-Nambu-Goldstone bosons. Moreover, by some rather straightforward extension of
the model we have found that a realistic pattern for the SM fermion masses and mixings emerges
naturally, without any flavor symmetry, from the most general interactions allowed by the gauge
symmetry and consistent with the geometric setup. 

It is worth noting that our model has good agreement with the MSSM for small values of $\tan\beta$.
One might argue that such values are strongly disfavored by the LEP limit, $\tan\beta>2.4$~\cite{LEP}.
However, on the one hand, the bound can be evaded if the stop mass and the relevant $A$-terms are large
enough~\cite{OK}. On the other hand, very small $\tan\beta$'s are still allowed in the (non-standard)
hidden Higgs scenarios~\cite{HH}.

We can assume that our model is embedded in some minimal supergravity theory, and that SUSY
breaking takes place through gravity mediation. It is only after SUSY breaking that the tree-level
vacuum degeneracy is resolved and a particular vacuum is picked up through radiative corrections.
It would be interesting to see how radiative corrections lift the flat directions so as to give
rise to a stable (local) minimum with the VEVs having a small component in a direction that breaks
$G_\text{SM}$ down to $SU(3)_c\otimes U(1)_\text{em}$. The circumstances under which this can happen 
were investigated in~\cite{bdm}. Such considerations would provide with phenomenological constraints 
of the model.

Finally, we briefly discuss proton stability in our model. As is known, it is the exchange of the
heavy colored triplets that dominates the contribution to possible proton decay. In a model like
ours such contributions are naturally suppressed~\cite{bdsbh}, as we will argue. First we note
that the triplets appear as Goldstone bosons \emph{only} in the breaking: $SU(6)\rightarrow SU(5)$,
which takes place because of the VEVs $\langle\Psi\rangle$ and $\langle\bar\Phi_i\rangle$'s.
Therefore, only the triplet coming from $\Sigma$ is physical, which remains as a heavy state,
and can potentially mediate proton decay. However, one can compute the mass matrix for
the Higgs and gaugino triplets to see that there is no mass mixing between the triplets coming
from $\Sigma$ and ($\Psi, \bar\Phi_i$). This renders the triplet coupling ineffective to proton
decay whenever the light Higgs doublet (that gives the relevant mass term) does not originate
from $\Sigma$\,.

\subsection*{Acknowledgments}

We would like to thank R. Barbieri, H.-C. Cheng, G. Dvali, and R. Rattazzi for useful comments.
We are especially thankful to H. Zhang for providing us with the reference values of fermion masses
at the GUT scale. RR was a Postdoctoral Fellow of the Fonds de la Recherche Scientifique-FNRS, whose work was partially supported by IISN-Belgium (conventions 4.4511.06 and 4.4514.08). RR was also partially supported by the ``Communaut\'e Fran\c{c}aise de Belgique" through the ARC program and by the ERC Advanced Grant ``SyDuGraM." The work of RR was also supported partially by ERC Grant n.226455 Supersymmetry, Quantum Gravity and Gauge Fields (Superfields), and by James Arthur Graduate Fellowship at New York University. RR is grateful to INFN Pisa, Scuola Normale Superiore and Universit\'e Libre de Bruxelles for hospitality. A.E.C.H was supported by Fondecyt (Chile), Grant No. 11130115 and by DGIP internal Grant No. 111458.

\end{document}